\documentclass{article}

\usepackage{arxiv}

\usepackage[utf8]{inputenc} 
\usepackage[T1]{fontenc}    
\usepackage{url}            
\usepackage{booktabs}       
\usepackage{amsfonts}       
\usepackage{nicefrac}       
\usepackage{microtype}      
\usepackage{lipsum}
\usepackage{amsmath}
\usepackage{graphicx}
\graphicspath{{images/}}
\usepackage{parskip}
\usepackage{fancyhdr}
\usepackage{bm}
\usepackage{witharrows}
\usepackage{float}
\usepackage{multirow}
\usepackage{caption}
\usepackage{booktabs}       
\usepackage[hidelinks]{hyperref}
\usepackage{multirow}
\usepackage[nameinlink]{cleveref}
\usepackage{amsmath} 

\title{Triplet Entropy Loss: Improving The Generalisation of Short Speech Language Identification Systems}

\author{Ruan van der Merwe \\
  Department of Statistical Sciences\\
  University of Cape Town\\
  South Africa\\
  \texttt{ruanh.vandermerwe@gmail.com} \\
}

\begin{document}
\maketitle

\begin{abstract}
We present several methods to improve the generalisation of language identification (LID) systems to new speakers and to new domains. These methods involve Spectral augmentation, where spectrograms are masked in the frequency or time bands during training and CNN architectures that are pre-trained on the Imagenet dataset. The paper also introduces the novel Triplet Entropy Loss training method, which involves training a network simultaneously using Cross Entropy and Triplet loss. It was found that all three methods improved the generalisation of the models, though not significantly. Even though the models trained using Triplet Entropy Loss showed a better understanding of the languages and higher accuracies, it appears as though the models still memorise word patterns present in the spectrograms rather than learning the finer nuances of a language. The research shows that Triplet Entropy Loss has great potential and should be investigated further, not only in language identification tasks but any classification task.
\end{abstract}


\section{Introduction}
Speech recognition tools have grown to form an integral part of many lives. For example, if you are reading this paper on an electronic device, the chances of it containing one or other speech recognition tool are high, with Siri and Alexa being widely known examples. To build a speech tool, one must first have a back-end that can perform Automatic Speech Recognition (ASR), which is a process to determine automatically what a user said purely based on the input speech signal. One of the first steps in an ASR system is to identify the spoken language using language identification (LID) systems, as the ASR systems in most cases are optimised for one language only. When the spoken language is one of the more common languages in the world, then there are already a vast majority of resources and tools that could implement this identification step. But it is becoming more apparent that some of these tools do not generalise well to new domains \cite{abdullah2020crossdomain}. There also are cases where the performance degrades when new speakers are introduced \cite{montavon2009deep}. 

We present and investigate methods that can improve the generalisation to new domains for LID systems, specifically in the case of short speech utterance (three seconds). The techniques will be tested on South African languages in order to test the performance on low resource languages as well if the techniques work on languages that share the same characteristics, such as \textit{Zulu} and \textit{Xhosa}.  The new LID techniques could then improve the ASR systems in South Africa. 

Our main contribution is a novel training method called Triplet Entropy Loss (TEL). The TEL training method consists of training Convolutional Neural Networks (CNN) by optimising a loss function that combines the strength of Cross Entropy Loss (CEL) and Triplet loss. The TEL method aims to improve upon the work done by \cite{abdullah2020crossdomain} in terms of domain generalisation. Their approach involves applying domain adaption through adversarial training, which resulted in better generalisation  for Slavic languages. For their, technique one still requires data from the different domain during the training phase while TEL only requires one domain for training. The training domain is the NCHLT speech corpus \cite{barnard2014nchlt} and the unseen domain will be the Lwazi speech corpus \cite{inproceedings} .

We will start by first giving an overview of the work that has been done in LID systems, whereafter we will go over the TEL training method as well as the other techniques mentioned. We will then focus on the results obtained on the NCHLT and Lwazi domains.

\section{Previous Work}\label{previous_work}

The first attempt of creating a system to predict the spoken language for all eleven official South African languages was by \cite{davel2012validating}, where they implemented a Parallel Phoneme Recognition followed by Language Modeling (PPR-LM) architecture. For the PPR-LM system, the researchers built a phoneme extractor for all eleven languages using Hidden Markov Models (HMM). For every audio input signal, they sent the signal through the HMM to extract the phonemes whereafter the biphone frequencies are extracted from the identified phoneme strings. They then concatenate these frequencies into a single vector where it is then fed into a Support Vector Machine (SVM) to predict what the spoken language is. They trained this system on the \textit{Lwazi} speech corpus and produced fairly good results. With audio inputs ranging from 3 to 10 seconds they achieved an accuracy of $71.72\%$, but this drastically increased when they tried to predict the family language of the audio input (\textit{Afrikaans}, \textit{English}, \textit{Nguni}, \textit{Sotho-Tswana}, \textit{Tswa-Ronga} or \textit{Venda}). Similar techniques were also used by \cite{henselmans2013phoneme}. 

In 2017, \cite{bartz2017language} identified a way to use a hybrid model built out of a CNN and a LSTM. What separated their work from previous work was that they built a CNN which received the image of a raw audio signal converted to a spectrogram as the input. The researchers then fed the extracted features of the CNN into an LSTM to generate the predictions. Similar methods were applied in \cite{revay2019multiclass} and \cite{sarthak2019spoken}. The researchers in \cite{margolisapplying}, \cite{bredin2017tristounet} and \cite{mingote2019language} looked at using Triplet loss on various speech tasks such as LID and user identification. \cite{mingote2019language} specifically looked at implementing a LID system based on triplet networks. Their method alters the Triplet loss function to optimise the Area Under the Curve (AUC). The researchers have also previously implemented this same technique for text-dependent speaker verification systems \cite{mingote2019optimization}. By implementing this loss, the researchers managed to outperform traditional methods, such as Weighted Gaussian Back-end and SVM's, in the context of a closed-set evaluation of the LRE 2009 dataset, \cite{mingote2019language}.

A recent study from \cite{abdullah2020crossdomain} has found that end to end deep learning LID systems perform severely worse when tested on out of domain samples. The researchers define out of domain samples as data coming from datasets the model was not trained on. To improve the generalisation from domain to domain, the researchers aimed to force the model to learn features that are domain invariant. They do this by using a technique called domain adaption through backpropagation that was introduced by \cite{ganin2014unsupervised} and also mentioned in \cite{meng2017unsupervised}. The technique entails creating two fully connected network blocks, where each block has its own task, that are connected to the output of a feature extraction block. The block $B_y$ aims to predict the language of a sample while the block $B_d$ tries to predict from which domain a sample is sampled. Both these blocks get fed by $B_f$, which is the feature extractor. Each training sample in the source domain gets augmented with a label $d=0$ and the samples from the other domain receive a label $d=1$ but the samples do not have a label showing the language. The parameters $\theta_d$, which are the parameters of the block predicting the domain, are then optimised to only minimise the loss of the domain classifier and likewise $\theta_y$ is optimised to predict the language label.

To ensure that the parameters of the feature extractor block become uninformative of the domain, the researchers seek to optimise $\theta_f$ such that the loss of the domain classifier gets maximised. This is known as adversarial training, where there is a competition between blocks in the network to optimise different losses. The loss function for this method can be found in \Cref{cross_domain_loss} where $\mathcal{D}_{S}$ is the source domain, $\mathcal{D}_{\mathcal{T}}$ is the target domain and $\lambda$ is a parameter that controls the contribution of the domain classifier’s loss \cite{abdullah2020crossdomain}.

\begin{equation}
\begin{array}{c}
J\left(\boldsymbol{\theta}_{f}, \boldsymbol{\theta}_{y}, \boldsymbol{\theta}_{d}\right)=\sum_{\left(\mathbf{X}_{i}, y_{i}\right) \in \mathcal{D}_{S}} L_{y}\left(B_{y}\left(B_{f}\left(\mathbf{X}_{i} ; \boldsymbol{\theta}_{f}\right) ; \boldsymbol{\theta}_{y}\right), y_{i}\right) \\
-\lambda \sum_{\left(\mathbf{X}_{t}, d_{i}\right) \in\left(\mathcal{D}_{\mathcal{S}} \cup \mathcal{D}_{\mathcal{T}}\right)} L_{d}\left(B_{d}\left(B_{f}\left(\mathbf{X}_{i} ; \boldsymbol{\theta}_{f}\right) ; \boldsymbol{\theta}_{d}\right), d_{i}\right)
\end{array}
\label{cross_domain_loss}
\end{equation}

The researchers tested this method with the Slavic portion of the GlobalPhone Read Speech (GRS) dataset as the source domain and the Radio Broadcast Speech (RBS) dataset as the other domain. The RBS dataset is a large collection of Slavic radio broadcasts. The researchers also used CNNs as base models. When not applying the domain adaption technique described above, the models performed very well in domain but performed poorly for out of domain samples. After applying the cross domain adaption technique, the researchers could achieve improvements up to $77.7\%$.

In this work we aim to design techniques that allow an LID system to generalise to new domains without requiring data from the new domain. This is accomplished by the TEL training method discussed in the next section, as well as using Spectral Augmentation and pre-trained CNN architectures. Spectral augmentation will be applied directly to the spectrograms during the training phase, as was done by \cite{park2019specaugment}. Spectral augmentation comprises masking the spectrogram image horizontally or vertically and sometimes both directions. 

\cite{palanisamy2020rethinking} showed that CNN architectures pre-trained on the Imagenet dataset serve as a good baseline for training audio classification models. In order to test how well this translates to the prediction of South African languages, the experiments performed will also feature the results for when the models are pre-trained. Even though \cite{palanisamy2020rethinking} experimented with which layers to freeze during the training process, this work will allow all layers in the model to be trainable during the training process. This is done to reduce the amount of experiment runs that need to happen. It is then assumed that this will not affect the results greatly, but is left as future work to investigate further.

\section{Triplet Entropy Loss}\label{methodlogogy}

In multi-class classification networks, the de facto technique to train a network is to generate one-hot encoded vectors from the class labels and use Cross Entropy Loss (CEL) to calculate the respective loss value. The formula for CEL is shown in \Cref{ce_function} where $y_{i}^{c}$ is the probability that the $i$th observation belongs to class $c$, with $\hat{y}_{i}^{c}$ being the predicted probability of the $i$th observation belonging to class $c$. By optimising the network using \Cref{ce_function}, the weights are optimised in such a way as to improve the estimated probability distribution of the network, such that the correct class has the highest probability output given the input features $X$.

\begin{equation}
CEL=-\sum_{i=1}^{N}\sum_{c=1}^{C}y_{i}^{c}\log{\hat{y}_{i}^{c}}
\label{ce_function}
\end{equation}

The downside with this loss function is that it only penalises the output of the class under consideration as $y_{i}^{c}$ is zero for all cases where $y_{i}^{c}$ is not the target class. Even if training a model using mini-batch gradient descent, where the weights are optimised based on the average loss of the batch which contains multiple classes, the individual losses used to calculate the average still only considers the correct class prediction and not the overall class interaction for that pass through. Even if the Softmax function is used as a predecessor to the CEL, where the output is calculated by looking at all the class outputs, the final value used in CEL is still just the Softmax output for the given class which is not a reflection of the interactions between the classes for that specific prediction.

For tasks such as language identification where the input data that is present in various classes, such as someone speaking a mix of words from different languages, it will be more optimal to have a loss function that interprets interactions between classes. The loss must optimise the network by learning these interactions between classes to generalise better to the instances where there is a tiny threshold between the classes. A loss that loosely fits this description is the Triplet loss function used in \cite{margolisapplying}, \cite{bredin2017tristounet} and \cite{mingote2019language} for language identification tasks. By using the Triplet loss function, the weights are being optimised by comparing different class embeddings with one another and optimising the distance between the embeddings such that different classes are far from one another. The model can then learn special characteristics of all the classes and in doing so could be able to better learn the characteristics between languages. This is all well, but Triplet loss does not optimise for prediction capabilities directly.

\begin{figure}[H]
    \centering
    \includegraphics[scale=0.4]{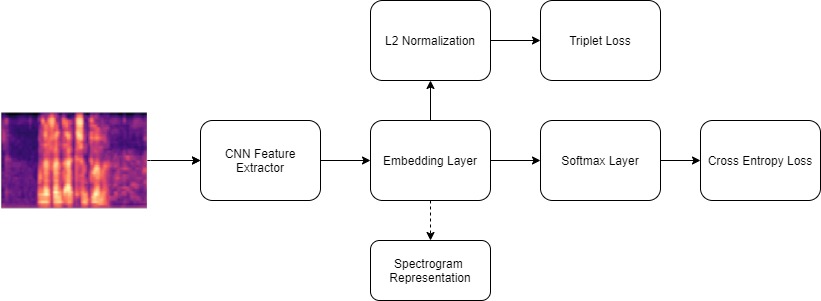}
    \caption{Triplet Entropy Loss high level overview}
    \label{figure_of_training_procudure}
\end{figure}

We present the novel training method Triplet Entropy Loss (TEL) that leverages both the strengths of CEL and Triplet loss during the training process, assuming that it would lead to better generalisation  for language identification tasks. \cite{khosla2020supervised} implemented something very similar for image classification, as they pre-trained a network using supervised Contrastive loss, whereafter they fine tune the model for classification tasks. The TEL method does not contain a pre-training step, but trains simultaneously with both CEL and Triplet loss, as shown in \Cref{figure_of_training_procudure}. As seen, the final embedding layer feeds into two separate layers where each of these output layers are connected to two different losses. TEL can be represented by \Cref{tel_formula}, with $\sigma$ being the Softmax function, $f()$ the embedding function and $g()$ the final classification layer. $N$ is the number of examples in the batch being passed through the network. The embeddings generated for the anchor, positive and negative triplets are given by $f(x_{i}^{a})$, $f(x_{i}^{+})$ and $f(x_{i}^{-})$. If only CEL is used, $f(x_{i}^{a})$ will be the vector passed to the Softmax layer. 

\begin{equation} \label{tel_formula}
\begin{split}
TEL 
&= \sum_{i=1}^{N} CEL_{i} + TL_{i}\\
& = \sum_{i=1}^{N}-y_{i}^a\log{\sigma(g(f(x_{i}^{a}))} +\left[\left\|f(x_{i}^{a})-f(x_{i}^{+})\right\|_{2}^{2}-\left\|f(x_{i}^{a})-f(x_{i}^{-})\right\|_{2}^{2}+\alpha\right]_+
\end{split}
\end{equation}

Looking at \Cref{tel_formula}, it can be seen that the model is being optimised to generate embeddings for $f(x_{i}^{a})$ that are strong at predicting the correct class while still ensuring that the interactions between other classes are not ignored. For example, if the weights for the given batch produce high probabilities for the correct class, but do not separate the classes well in the embedding space, the model will still be penalised to further separate classes. In doing so the model will learn what are the features that truly distinguish the classes. This will make the model more effective in new domains because by increasing the distance between classes, it is assumed that less precise predictions made on new domain data will still have room for error and still produce well separated embeddings.

During training of the network, it is very important that triplets are chosen in such a manner that they are “hard” for the network. In other words, if the only triplets being chosen are instances where $\left\|f(x_{i}^{a})-f(x_{i}^{+})\right\|+\alpha < \left\|f(x_{i}^{a})-f(x_{i}^{-})\right\|$, then the network will learn nothing as the contribution to the loss (from the triplets) will be $0$. To ensure this does not happen, the triplets generated and fed into the loss will be mined in each batch such that $\left\|f(x_{i}^{a})-f(x_{i}^{-})\right\| < \left\|f(x_{i}^{a})-f(x_{i}^{+})\right\|$. This is known as semi-hard negative online mining and was found to lead to better local minima discovery and help in reducing the chance that the model collapses and predicts all embeddings to be $f(x) = 0$ \cite{schroff2015facenet}.

It is also assumed that by using TEL the embedding layer is optimised in such a way that the generated embeddings will give an accurate description of the language spoken, especially in cases where it is not someone’s first language. For example, speakers that mix their languages frequently, such as \textit{Afrikaans} speakers using a lot of \textit{English} words, should appear closer to one another than speakers that that only use \textit{Afrikaans} words. Another example could be that a famous \textit{Afrikaans} news host’s embeddings should not be close to a speaker learning how to speak \textit{Afrikaans}, but both should still form part of the cluster for \textit{Afrikaans} speakers. 

Using TEL, predictions can be made with a model trained on the generated embeddings or by using the same technique such as \cite{schroff2015facenet} where the distance between embeddings can be used with a voting system implemented. Future work could look into pre-training a network using TEL and then fine tuning the model for classification tasks. To see the performance of the novel TEL method, please refer to \Cref{nchlt_results} and \Cref{lawzi}.

\section{Results on the NCHLT Dataset}\label{nchlt_results}

In order to test if the TEL method truly results in better performance, three models are trained in three different ways. The three models are the CRNN model used in \cite{bartz2017language}, a Resnet-50 model and a Densenet-121 model. All three  models are trained using the TEL method as well as CEL and Triplet loss separately. Because of the constraints in compute resources and time, there will be no cross validation performed. It is assumed that the out of training sample datasets contain enough variation to test the validity of the model performance.

The models trained without spectral augmentation will serve as a baseline for that respective architecture and training method. The success of these experiments are measured with the accuracy metric and by observing the performance for each language separately by inspecting the confusion matrix. With the confusion matrix, the confusion that appears between languages can also be observed. Along with this, the embeddings produced will be inspected visually. Performance is measured across all samples as well as for each gender. This is to ensure the model is not biased towards any gender. The gender associated with each recording is taken directly from the metadata provided in the NCHLT dataset

The models are trained using the Adam optimiser \cite{kingma2014adam} with a learning rate of 0.0001. To reduce overfit within the models, the feed forward layers in all the models as well as the convolutional layers in the CRNN model contain an L2 penalty. All models also use a batch size of 32, as this is the maximum size the available VRAM can handle. The embedding layer produces a 512 dimensional vector in all models and the input spectrograms are three seconds in length and consist of 128 mel frequency bins.

\begin{table}[H]
\centering
\caption{Summary of the experiment results regarding the models trained using only Cross Entropy Loss (CEL) and Triplet Entropy Loss (TEL). The results in the table show the accuracy ($\%$) of the models on unseen NCHLT speaker data.}
\begin{tabular}{ccccc}
\hline
\textbf{\textit{Model}} & \textbf{\textit{CEL Baseline}}  & \textbf{\textit{CEL}}  & \textbf{\textit{TEL Baseline}} & \textbf{\textit{TEL}} \\ \hline
CRNN  & 71 & 70 & \textbf{75} & 74  \\
Densenet121 (Imagenet) & 78 & 78 & 79 & \textbf{81} \\
Densenet121  & 75 & 74 & 75 & \textbf{77} \\
Resnet50 (Imagenet) & 76 & \textbf{78} & 77 & \textbf{78}\\
Resnet50  & 74 & 72 & 76 & \textbf{78}\\ \hline
\end{tabular}
\label{exp_results_ce}
\end{table}

In \Cref{exp_results_ce} one can find the various accuracies produced by the TEL and CEL trained models. The highlighted values show the highest accuracy for that specific model architecture. All the models’ best results come from the TEL training methods, except for Resnet50 model pre-trained on ImageNet data, even though the TEL method still produced the same accuracy as the CEL trained model. When looking at \Cref{exp_results_tripelt}, which shows the Triplet loss values, one can see that on average the languages are better separated when the model is trained using only Triplet loss. If one takes the embeddings generated by these models and predict the language in the same way as the FaceNet researchers, namely assigning the embedding the same language as its closest neighbour, then the top accuracy achieved is $78\%$. This is still lower than the accuracies produced using the TEL method.

\begin{table}[H]
\centering
\caption{Summary of the experiment results with regards to the models trained using only Triplet loss and Triplet Entropy Loss (TEL). The results in the table shows the loss value for the models on unseen NCHLT speaker data.}
\begin{tabular}{ccccc}
\hline
\textbf{\textit{Model}} & \textbf{\textit{Triplet Baseline}}  & \textbf{\textit{Triplet}}  & \textbf{\textit{TEL Baseline}} & \textbf{\textit{TEL}} \\ \hline
CRNN  & 0.220 & 0.150 & \textbf{0.095} & 0.096  \\
Densenet121 (Imagenet) & \textbf{0.068} & 0.077 & 0.076 & 0.070 \\
Densenet121  & \textbf{0.074} & 0.075 & 0.088 & 0.080 \\
Resnet50 (Imagenet) & 0.078 & \textbf{0.075} & 0.078 & 0.076\\
Resnet50  & 0.083 & 0.082 & 0.083 & \textbf{0.076}\\ \hline
\end{tabular}
\label{exp_results_tripelt}
\end{table}

Further, the bias of the models are inspected to see how well the models perform on males vs females. These results can be seen in \Cref{bias_val_nchlt}. All of the models perform roughly the same for both gender groups, with the Resnet50 models seeming to perform a better on males compared to females though.

\begin{table}[H]
\centering
\caption{Accuracy ($\%$) performance of the various models based on the gender of the speaker.}
\begin{tabular}{ccc}
\hline
\textbf{\textit{Model}} & \textbf{\textit{Female}}  & \textbf{\textit{Male}} \\ \hline
CRNN  & 76 & 76  \\
Densenet121 (Imagenet) & 80 & 81\\
Densenet121  & 75 & 76\\
Resnet50 (Imagenet) & 76 & 80\\
Resnet50  & 76 & 79\\ \hline
\end{tabular}
\label{bias_val_nchlt}
\end{table}

\begin{figure}
    \centering
    \includegraphics[scale=0.5]{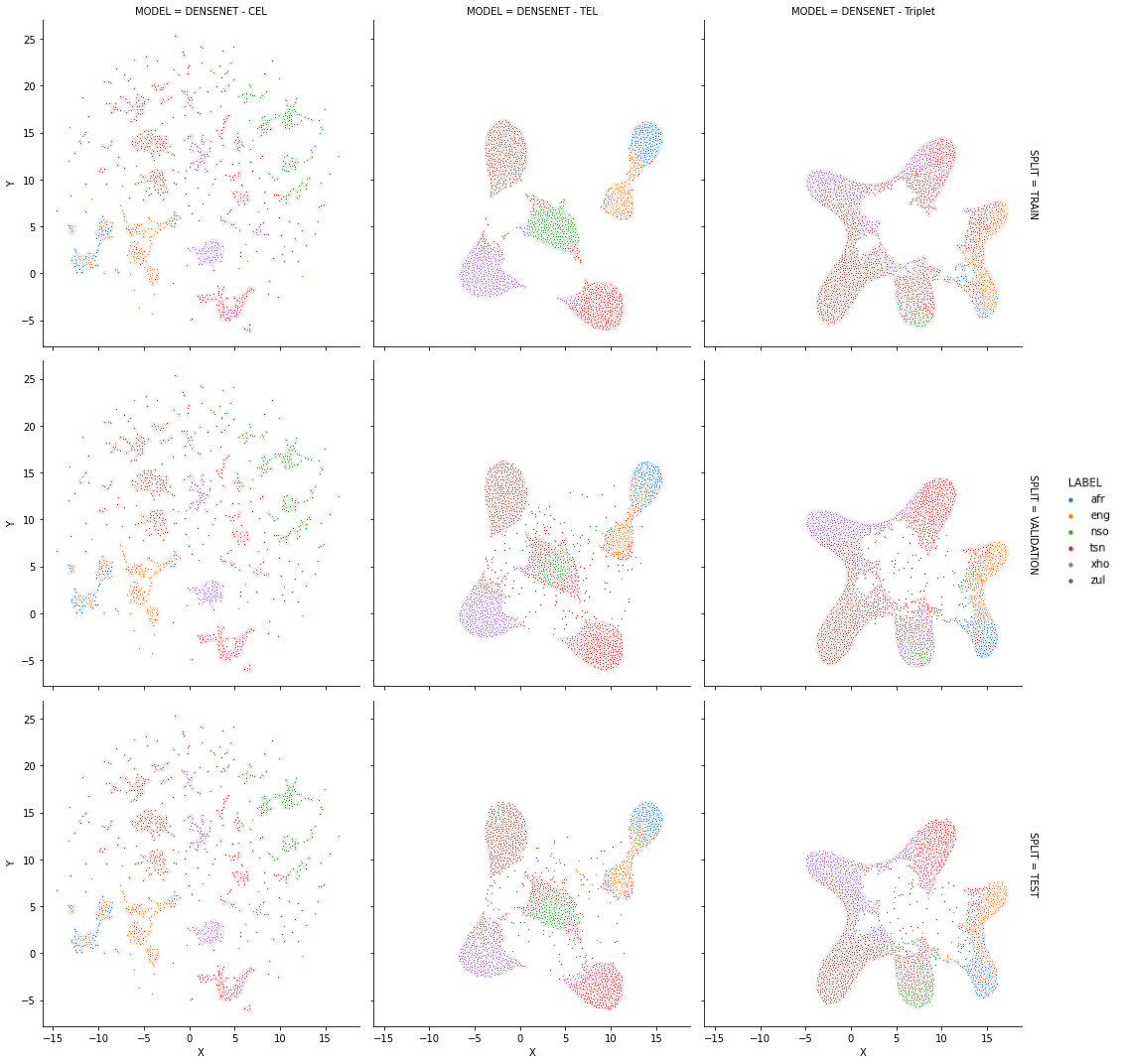}
    \caption{Projection of the embeddings generated by the Densenet121 model on the different NCHLT datasets.}
  \label{dense_projections}
\end{figure}

To further understand what the models learn during the different training methods, the generated embeddings are inspected visually. This is performed by training a UMAP model \cite{mcinnes2020umap} to project the embeddings down to two dimensions. UMAP is a manifold learning technique used for dimensionality reduction tasks. The model will look at the 15 closest neighbours when calculating the projections, with a minimum separation distance set at $0.001$, using the Cosine distance. The other hyperparameters are the default values set by the software package \cite{mcinnes2018umap-software}. The values were chosen in an iterative way until the clusters formed were interpretable. The UMAP model is trained using the NHCLT train set embeddings.

In \Cref{dense_projections} one can see the embeddings generated by Densenet121 model, pre-trained on Imagenet, for all of the training methods. One can see that embeddings created using the TEL method is better at separating the languages. This is clear if one looks at the training examples specifically, as the TEL embeddings create a cluster for all languages, but on the Triplet loss side \textit{English} and \textit{Zulu} is being confused with other languages.  Looking at the embeddings of the validation and test set embeddings, one can see that both models start confusing the languages, but that overall there are still clear language clusters in the TEL embeddings. The CEL trained model though has no clear clusters forming. The more interesting result come when one inspects the language families. Looking at the TEL embeddings, one can see that the Germanic languages (\textit{Afrikaans} and \textit{English}) form their own overlapping cluster to the right with the African languages forming their own clusters to the left. Looking at the validation and test embeddings, it can be seen that the confusion occurs more often between the families than with other languages.


\begin{figure}[H]
    \centering
    \includegraphics[scale=0.28]{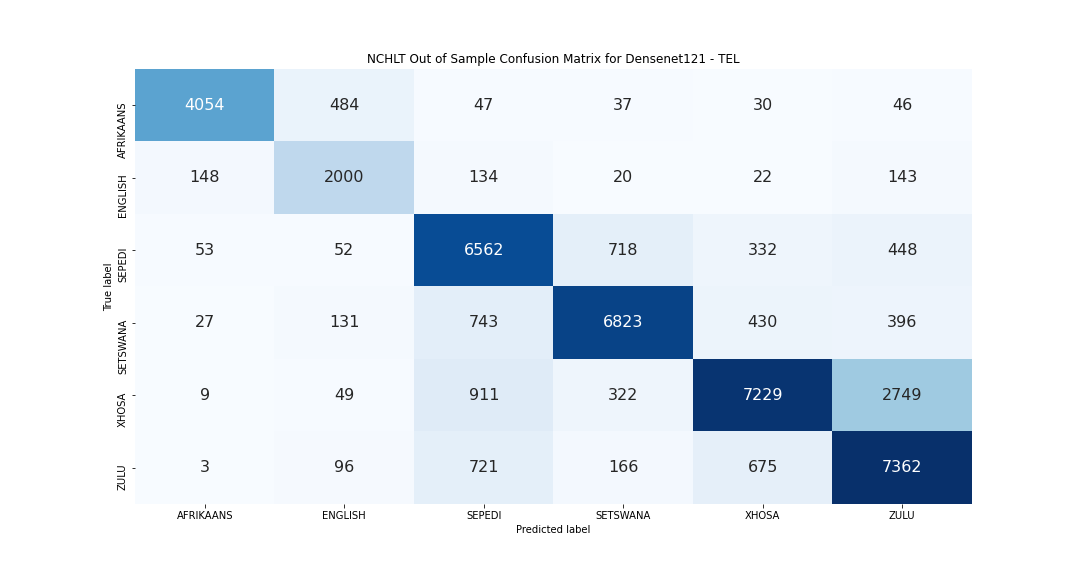}
    \caption{Confusion matrix generated by combining the validation and test set predictions for the pre-trained Densenet-121 model fine tuned using the TEL method.}
  \label{dense_nchlt_cm}
\end{figure}

This confusion appears less in the TEL model than the Triplet loss model, and this leads to the assumption that the TEL learning method is better at understanding the nuances of the languages. In the confusion matrix for the TEL Densenet121 model found in  \Cref{dense_nchlt_cm}, the confusion can clearly be seen (pun partially intended). The Germanic languages are not really confused with other languages but \textit{Xhosa} and \textit{Zulu} have the most confusion between them. Besides \textit{Zulu}, \textit{Sepedi} also gets confused with the other African languages. The confusion could be related to the words present in the prompts given to speakers that share the same words between languages.

The results also show that the techniques introduced in \cite{park2019specaugment} and \cite{palanisamy2020rethinking} can apply to South African language identification systems to improve the overall accuracy of LID systems. Spectral augmentation must however be investigated for each use case and each architecture. The reason for this is that the increase in accuracy could perhaps not be worth the extra complexity added to the system. Models trained with augmentation does however tend to create embeddings that separate the languages better than the respective baseline models and is therefore worth considering. 

\section{Results on the Lwazi Dataset}\label{lawzi}

In this section the Densenet121 and Resnet50 models that were trained on the NCHLT speech corpus are tested on out of domain samples to test how well the TEL training method generalises to a new domain. The recordings for the Lwazi dataset were sampled at $8kHz$ versus the NCHLT dataset which was sampled at $16kHz$. This will undoubtedly influence the results, as will the difference in words spoken between the two datasets.

\begin{table}[H]
\centering
\caption{Accuracy ($\%$) performance of the Resnet50 and Densenet121 models on the Lwazi dataset.}
\begin{tabular}{ccc}
\hline
\textbf{\textit{Model}} & \textbf{\textit{TEL Accuracy}}  & \textbf{\textit{CEL Accuracy}} \\ \hline
Densenet121 (Imagenet) & \textbf{45} & 25.8\\
Resnet50  & \textbf{28} & 23\\ \hline
\end{tabular}
\label{lwazi_accuracies}
\end{table}

\begin{figure}
    \centering
    \includegraphics[scale=0.4]{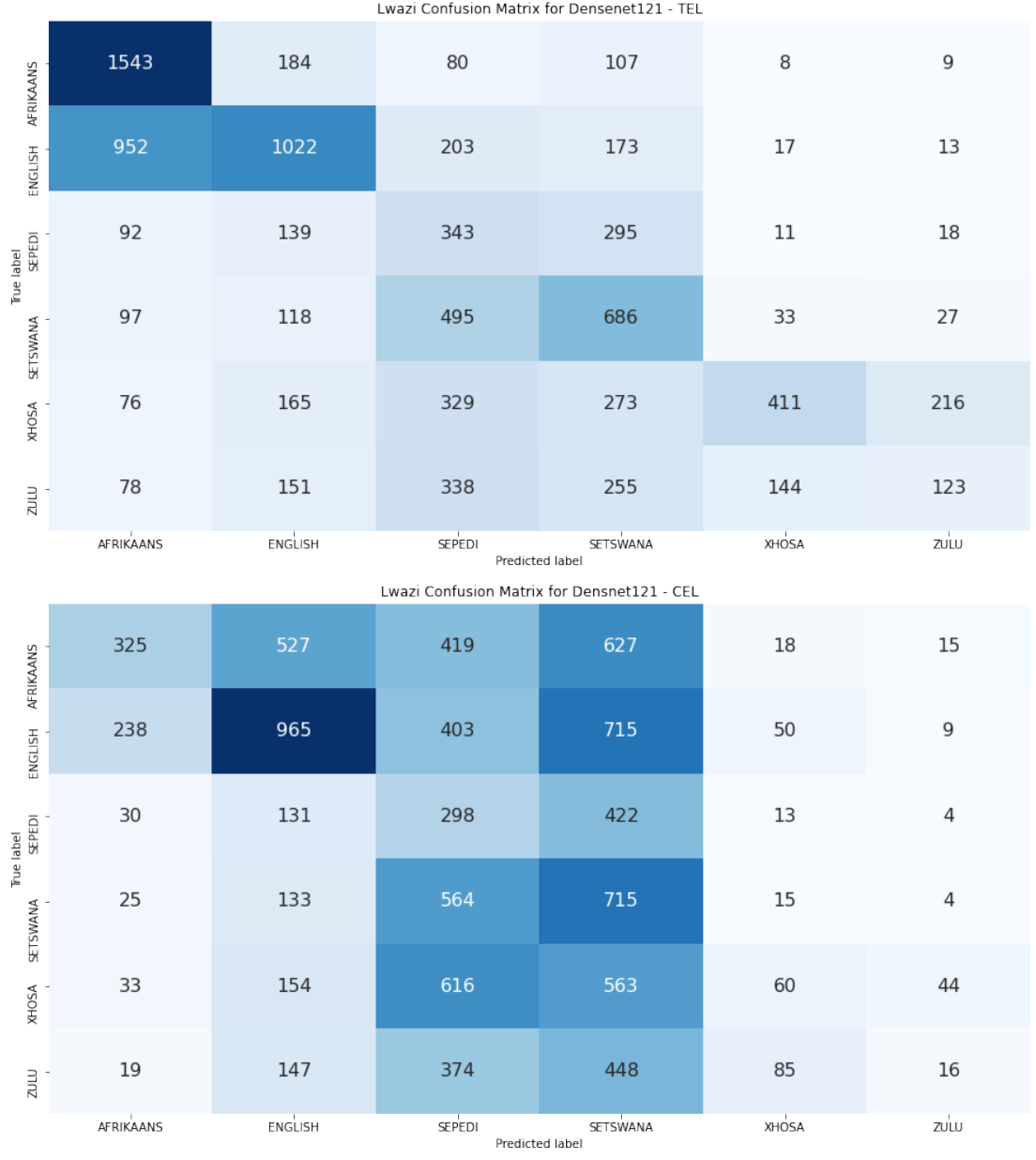}
    \caption{Confusion matrix's for the Densenet-121 models trained using the TEL method (top) and CEL (bottom).}
  \label{lwazi_cm_dense}
\end{figure}

In \Cref{lwazi_accuracies} one can find the performance of models on the new domain. As can clearly be seen, none of the models perform nearly as good as they did in the NCHLT domain. The only model that showed any promise is the Densenet121 model trained using the TEL method as it achieved close to $45\%$ accuracy. In both model cases, one can see that the TEL method does outperform the CEL method, but in the Resnet case it is by an ignorable margin. What the results do show though is that by combining the TEL method with the techniques in \cite{park2019specaugment} and \cite{palanisamy2020rethinking} produces a model that is more robust to new domains. This can be seen further in the confusion matrix for the Densenet121 models, found in \Cref{lwazi_cm_dense}. 

Looking at \Cref{lwazi_cm_dense} one will see that the CEL trained Densenet model tends to think that most languages are either \textit{Sepedi} or \textit{Setswana}, with very little samples being predicted to belong either to \textit{Xhosa} or \textit{Zulu}. The TEL trained model shows a better understanding for the Germanic languages as both \textit{Afrikaans} and \textit{English} have better results.  \textit{Xhosa} and \textit{Zulu} are also performing much better, showing a better separation in the African languages.

\section{Conclusion} \label{conclusion}

The research presented investigated several methods to improve the generalisation  of LID systems to new speakers and to new domains. These methods involve Spectral augmentation and using CNN architectures that are pre-trained on the Imagenet dataset. The primary method investigated though was the TEL training method which involves training a network simultaneously using Cross Entropy and Triplet loss. Several tests were run with three different CNN architectures to investigate what the effect all three methods have on the generalisation  of a LID system. The tests involved training models on six languages from the NCHLT speech corpus and measuring the performance of the models on new speakers from the same domain, and new speakers from a different domain.

It was found that all three of the methods do contribute to improve the accuracy of the models to the unseen data. The best two models were found to be a pre-trained Densenet121 model and a Resnet50 model trained from scratch, with both models being trained using the TEL method and Spectral augmentation. Even though both models could achieve $81\%$ and $78\%$ accuracies on new speakers in the NCHLT domain, both models still overfitted on the training data, as they achieved close to $100\%$ accuracy on the training speakers. When the models are tested on the Lwazi domain, the Resnet50 model performed poorly and only achieved an accuracy of $28\%$, whereas the Densenet121 model could achieve an accuracy of $45\%$. 

If one looks at the embeddings generated by both of the models for all training methods, the results do indicate that the TEL method has a better understanding of the languages. This is because of the stronger clusters formed in the TEL embeddings and the fact that the confusion happens more often between language families, whereas in the CEL/Triplet case the confusion occurs more often between all languages.

The results do also show that all the models might still only be memorising words rather than learning language characteristics. This will require more research, as is also recommended in \Cref{future_work}. The TEL training method also significantly reduces the training time required for a model to converge, specifically on the use cases shown in \Cref{some_more_huh}, which were done to show the robustness of TEL to different tasks.

\subsection{Recommendations for Future Research}\label{future_work}

The work presented only investigates six of the eleven South African languages. Languages from the Tswa-Ronga and Venda families were also not used in the research. Further work should be done to investigate how well these methods will work in systems where all eleven languages are present.

The model architectures and hyperparamaters presented in this paper are also not optimised yet. Future work, if there are better compute resources and a higher budget available, should further investigate all the parameters and hyperparameters used in this work.

Lastly, with the TEL method showing promise for audio classification tasks, it should also be investigated in other areas as well. Areas that will be a good fit for the TEL method are self-supervised tasks. Here Triplet loss could be substituted with Contrastive loss, forming Contrastive Entropy Loss. This is because there are various self supervised techniques out there already using Contrastive loss instead of Triplet loss and showing good results. The biggest challenges in these tasks will then be to generate labels to be used by CEL.

\section{Acknowledgements} \label{acknowledgements}
I would like to thank Dr Sebnem Er, my supervisor, for all the feedback and zoom calls and pushing me in the correct direction after every meeting and inspiring me to ask even more questions.

\newpage

\appendix

\section{Even More Use Cases}\label{some_more_huh}

To further test the TEL training method, the Densenet-121 and Resnet50 models will be trained on two different sound classification tasks. The first task is to predict the spoken digit in a recording, using the Free Spoken Digit Dataset (FSDD). The second task will be to predict which category of music a 30 second recording belongs to using the GTZAN dataset.

For both these tasks, the idea is to show how an out of the box model trained using the different methods (TEL, CEL and Triplet loss) performs on common datasets. The purpose of this is to further cement the hypothesis that TEL trained models do generate better embeddings than previous methods, even if the task is not language identification.

\subsection{Spoken Digits Classification}

The data used in this experiment is open source and can be found at \url{https://github.com/Jakobovski/free-spoken-digit-dataset/}. The dataset consists of six speakers, where each speaker utters a digit between zero and nine, fifty times each in \textit{English} only. This results in 3000 recordings. This can be imagined as a spoken MNIST dataset. 

\begin{figure}[H]
    \centering
    \includegraphics[width=\textwidth]{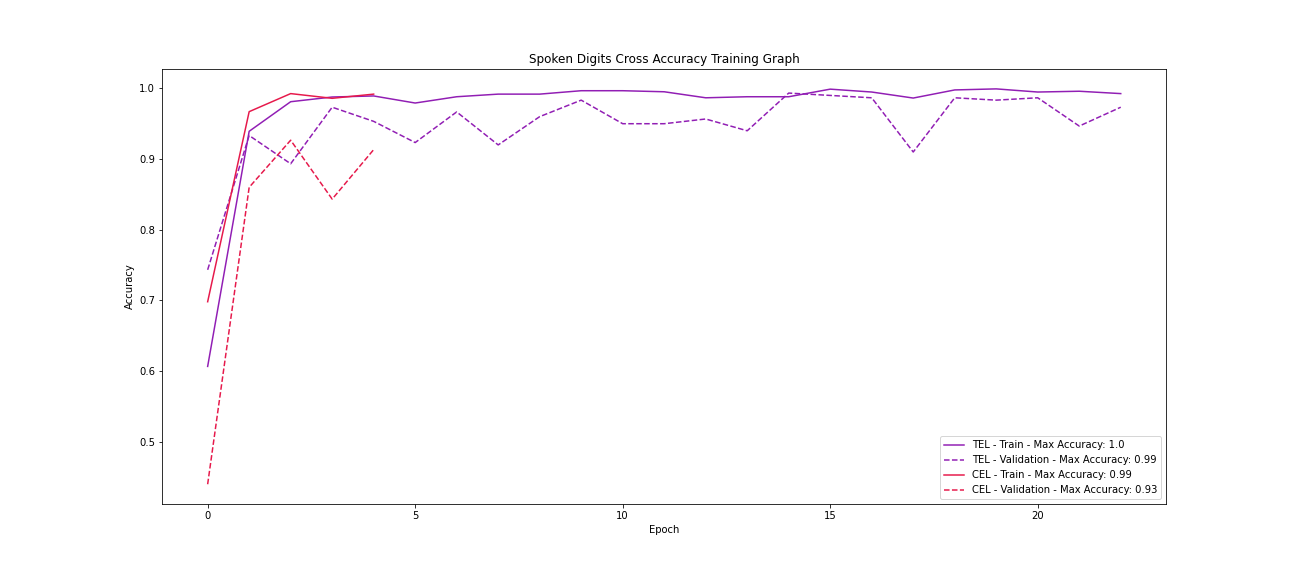}
    \caption{Training graph of pre-trained Densenet-121 model on the FSDD dataset in terms of accuracy.}
  \label{spoken_accuracy}
\end{figure}

The data is split into a train and validation set, with the validation set consisting of fifteen utterances for each speaker on each digit, totaling 2700 training samples and 300 validation samples. The models were again all trained using Adam with a learning rate of 0.0001 and a batch size of 64. Each model also had an early stopping criteria where if the validation loss did not improve for two epochs the training stops. In \Cref{spoken_accuracy} one can see the training graph in terms of accuracy for the pre-trained Densenet-121 models. As can be seen, both achieve a very high accuracy, but TEL has a higher accuracy and there is is smaller gap between the train and validation loss.

\begin{figure}
    \centering
    \includegraphics[width=\textwidth]{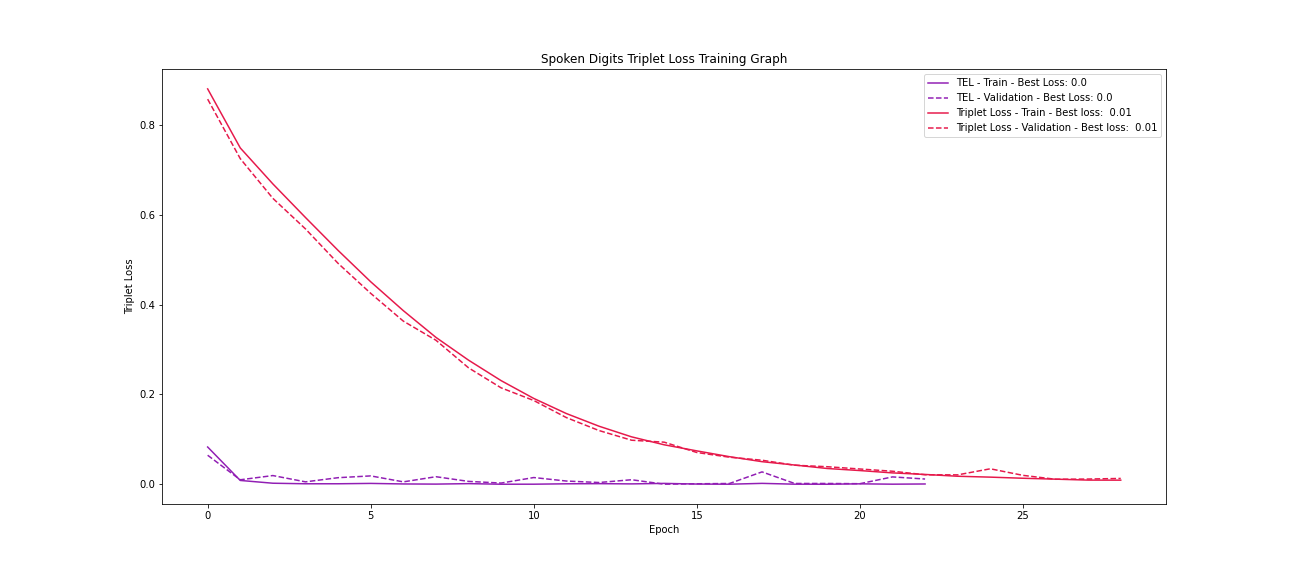}
    \caption{Training graph of pre-trained Densenet-121 model on the FSDD dataset in terms of Triplet loss.}
  \label{spoken_triplet}
\end{figure}

In \Cref{spoken_triplet} one can see the training graph in terms of the Triplet loss. Here one can see that both models again perform roughly the same, but the speed at which the TEL method reaches its minimum is much faster than the pure Triplet loss training method. This is the biggest differentiator between the two methods when applied to this dataset. 

\begin{figure}
    \centering
    \includegraphics[width=\textwidth]{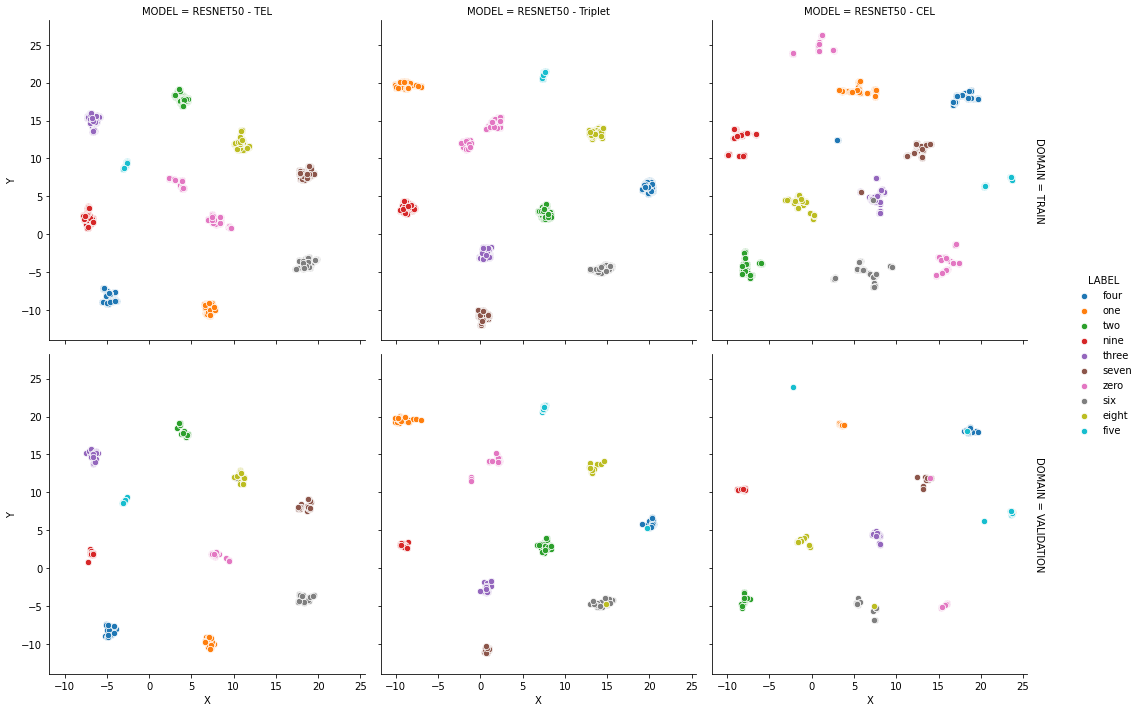}
    \caption{Projection of the embeddings generated on the FSDD dataset.}
  \label{spoken_emebddings}
\end{figure}

When looking at the embeddings generated (using UMAP again) in the \Cref{spoken_emebddings}, one will see there is not much difference here but it is important to note that the TEL method achieved these clusters after just the second epoch.

\newpage
\subsection{GTZAN - Music Genre Classification}

The GTZAN dataset contains 100 recordings for each of the ten distinct classes, namely blues, classical, country, disco, hip-hop, jazz, metal, pop, reggae, and rock. Each of the recordings are 30 seconds long. The version of the dataset used in this project can be found on Kaggle \href{https://www.kaggle.com/andradaolteanu/gtzan-dataset-music-genre-classification}{here}, where the audio is already converted to spectrograms for ease of use.

\begin{figure}[H]
    \centering
    \includegraphics[width=\textwidth]{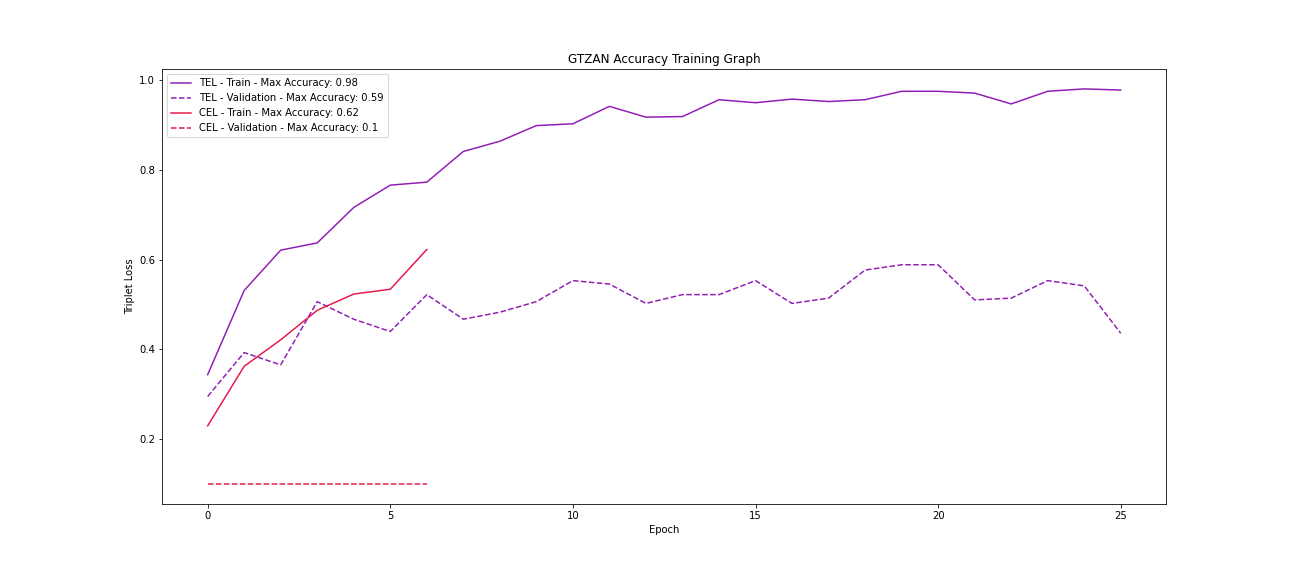}
    \caption{Training graph of pre-trained Densenet-121 model on the GTZAN dataset in terms of accuracy.}
  \label{gtzan_accuracy}
\end{figure}

The data is split into a train and validation set, with the validation set consisting of 21 recordings for each class. The exact model architecture and training process was followed as was done in the previous section with the FSDD dataset. In \Cref{gtzan_accuracy} one can see the training graph in terms of accuracy for the pre-trained Densenet-121 models. As can be seen, the validation accuracy for the model trained using CEL never increases while the validation accuracy for the TEL model reaches close to $60\%$ accuracy.

\begin{figure}[H]
    \centering
    \includegraphics[width=\textwidth]{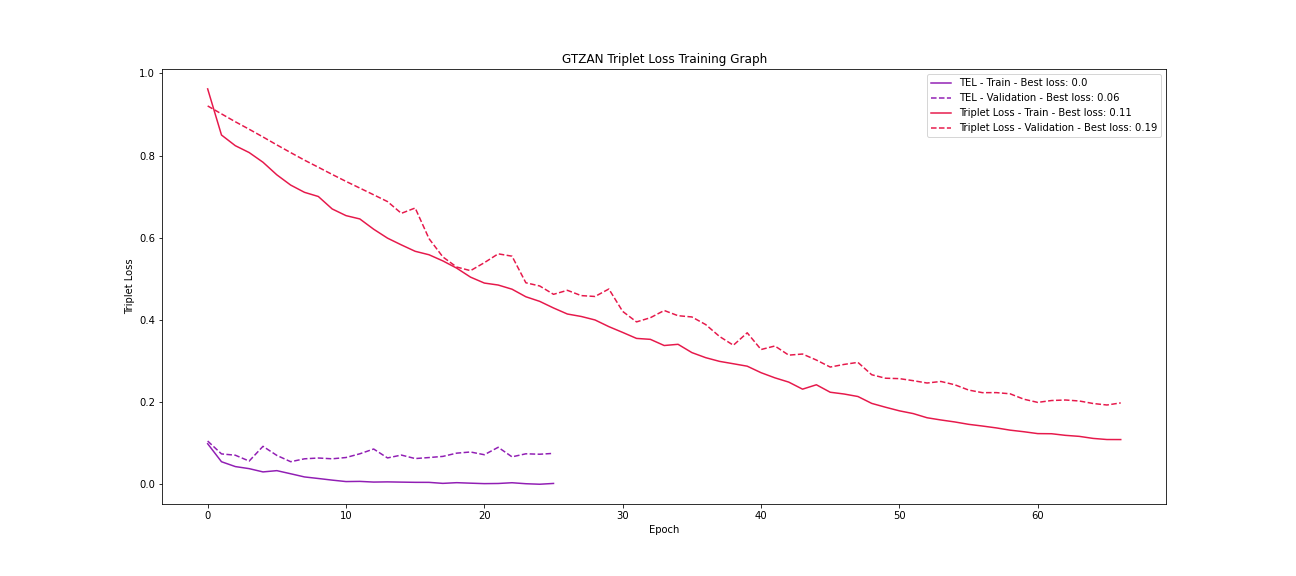}
    \caption{Training graph of pre-trained Densenet-121 model on the GTZAN dataset in terms of Triplet loss.}
  \label{gtz_triplet}
\end{figure}

In \Cref{gtz_triplet} one can see the training graph in terms of the Triplet loss. One again sees the same trend as in the previous section where the models, trained using TEL and Triplet loss, reach roughly the same performance but with the model trained using TEL reaching it much quicker. The TEL trained model though is able to separate the classes based on the Triplet loss value better than the other model.

\section{Projection of embeddings for other models}

\begin{figure}[H]
    \centering
    \includegraphics[scale=0.8]{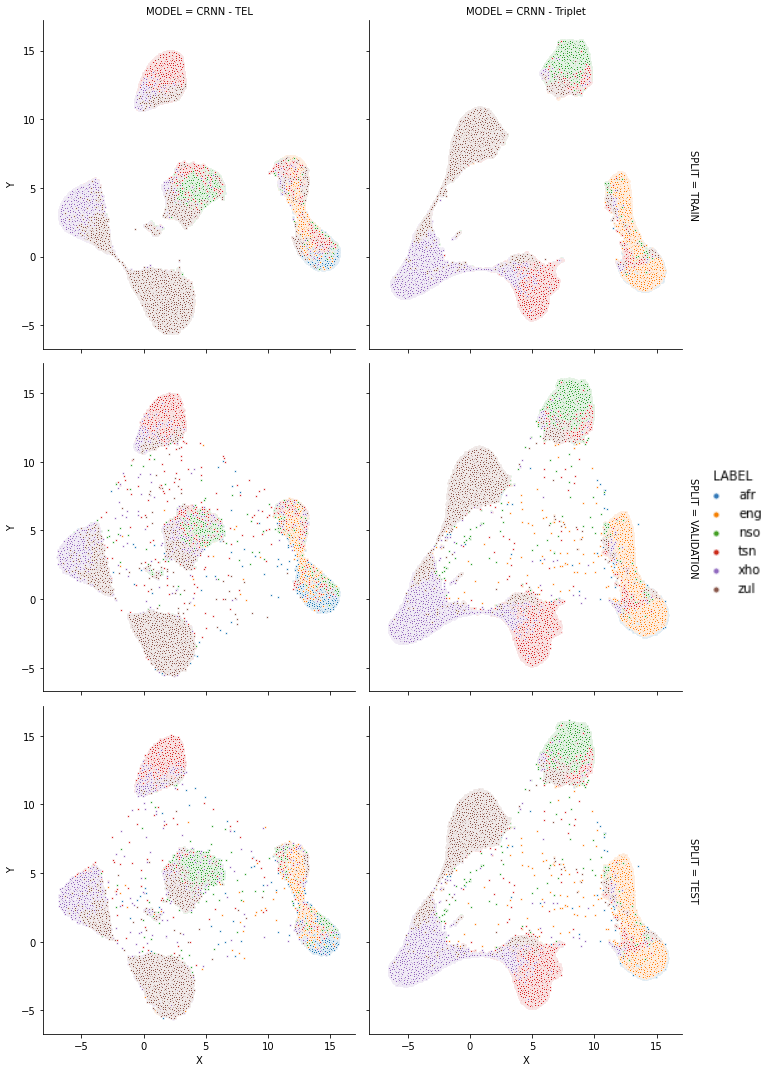}
    \caption{Embeddings generated by the baseline CRNN model on the NCHLT dataset.}
\end{figure}

\begin{figure}[H]
    \centering
    \includegraphics[scale=0.8]{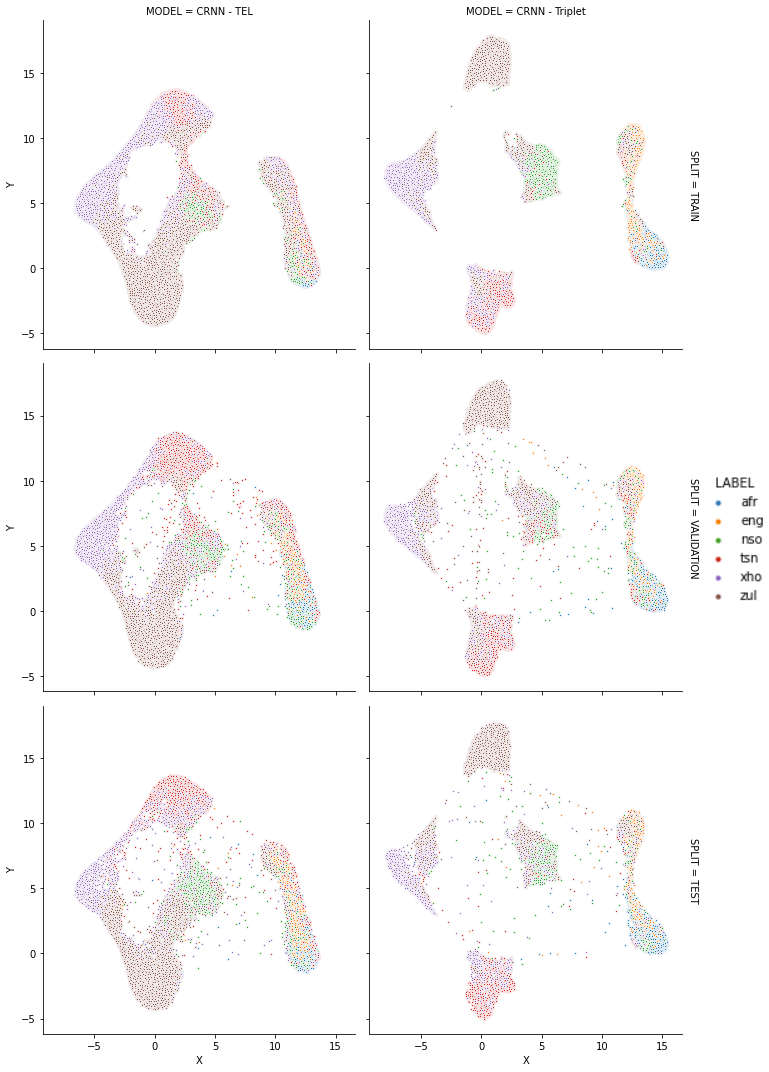}
    \caption{Embeddings generated by the CRNN model on the NCHLT dataset.}
\end{figure}

\begin{figure}[H]
    \centering
    \includegraphics[scale=0.8]{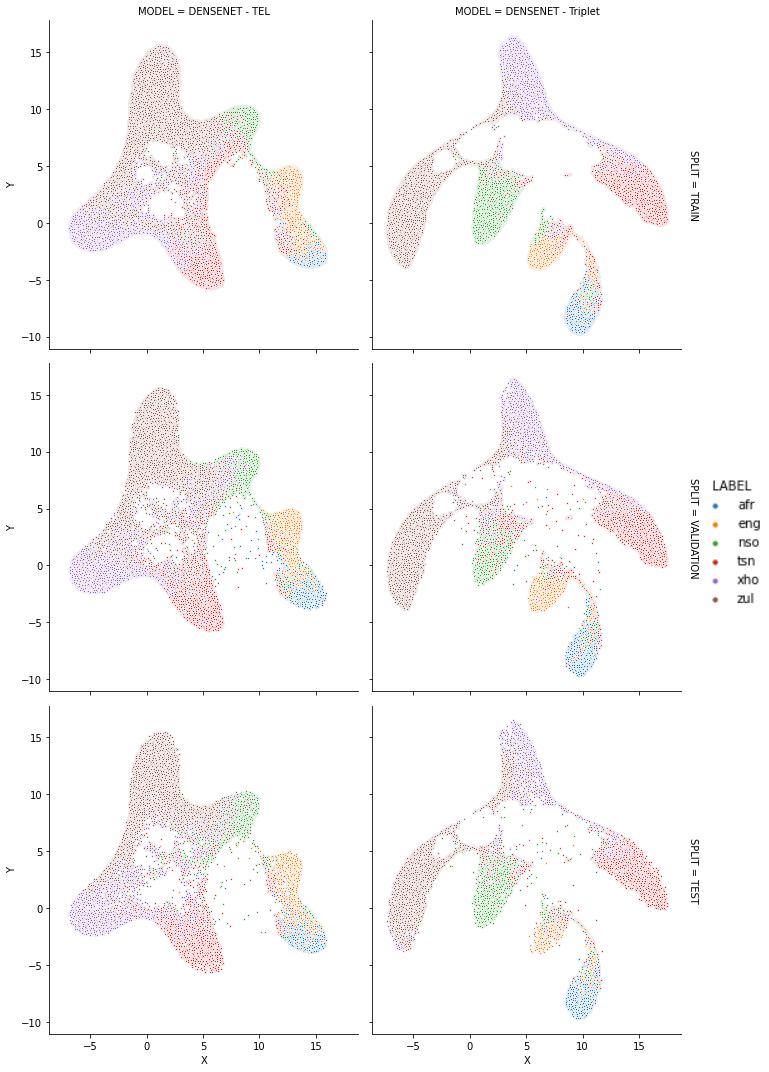}
    \caption{Embeddings generated by the baseline Densenet121 model on the NCHLT dataset.}
\end{figure}

\begin{figure}[H]
    \centering
    \includegraphics[scale=0.8]{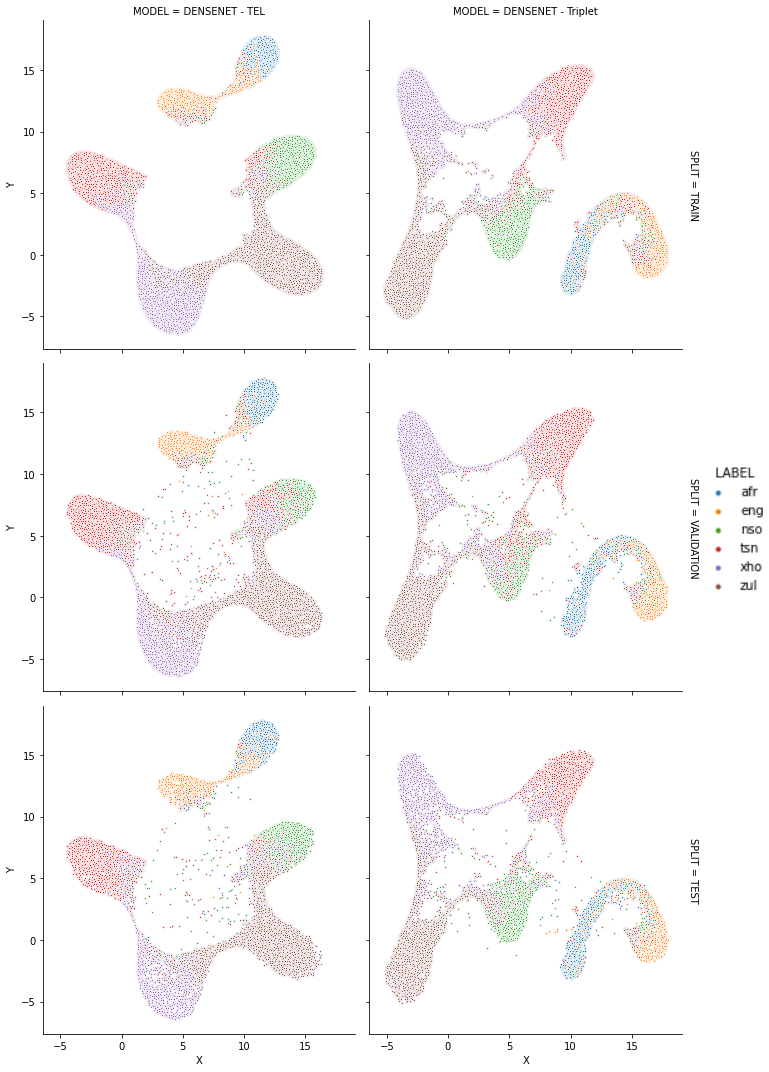}
    \caption{Embeddings generated by the Densenet121 model on the NCHLT dataset.}
\end{figure}

\begin{figure}[H]
    \centering
    \includegraphics[scale=0.8]{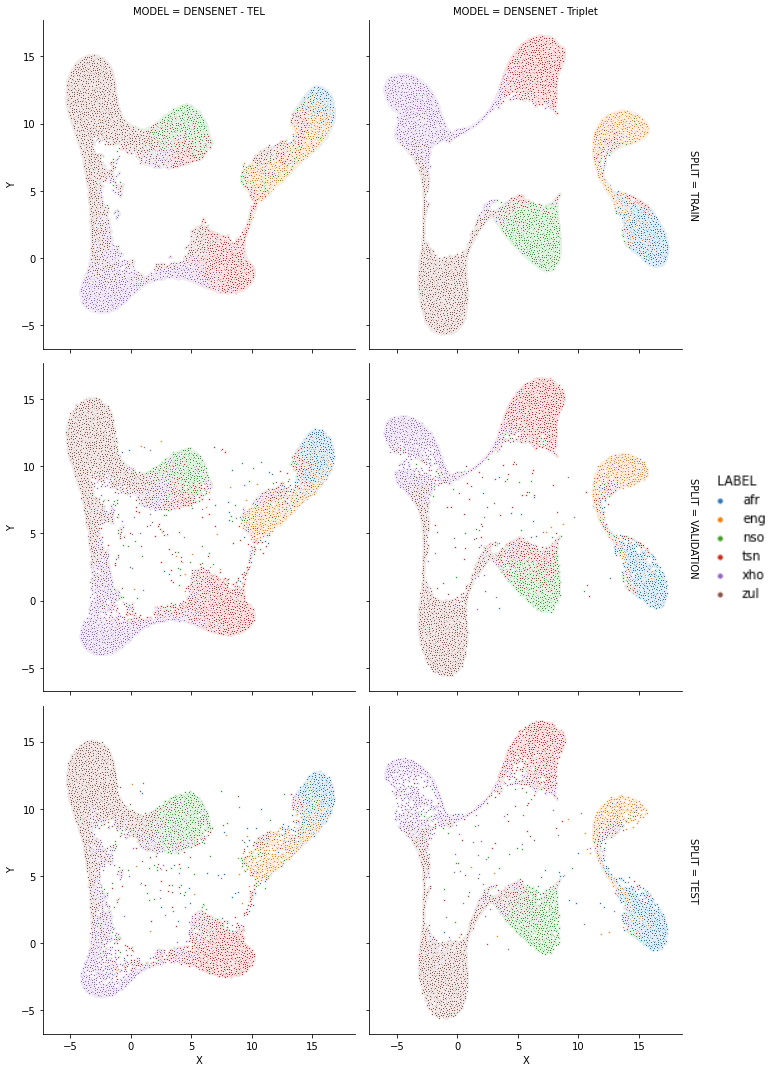}
    \caption{Embeddings generated by the baseline pre-trained Densenet121 model on the NCHLT dataset.}
\end{figure}

\begin{figure}[H]
    \centering
    \includegraphics[scale=0.8]{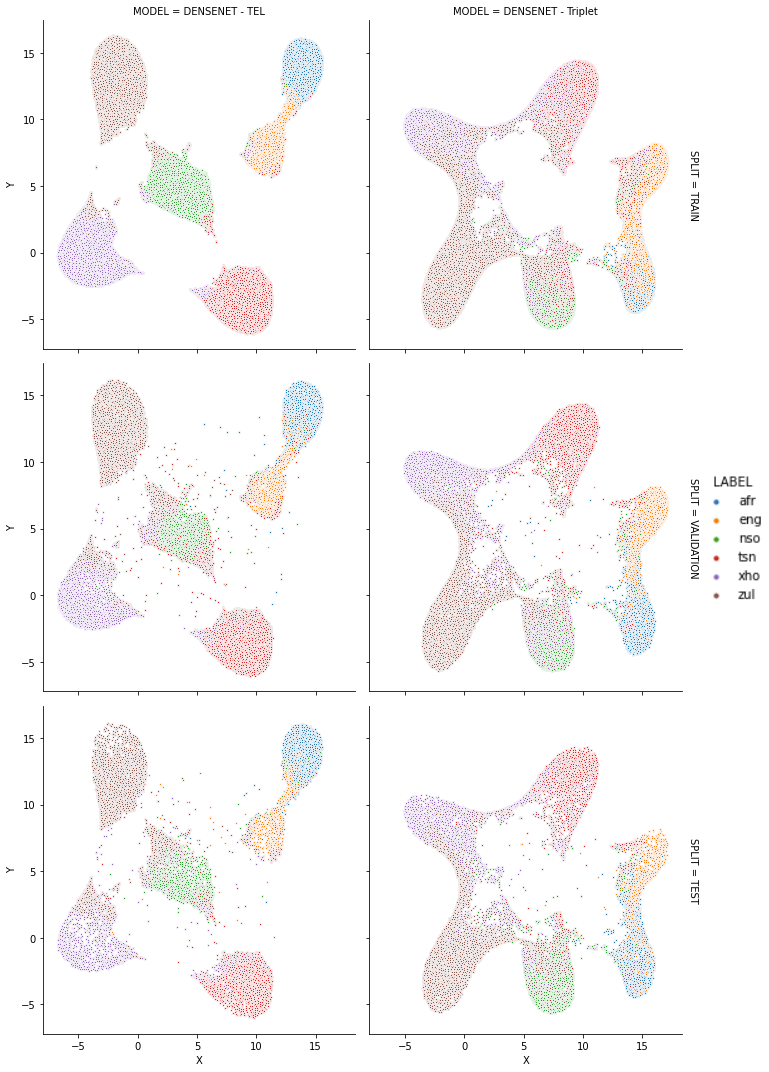}
    \caption{Embeddings generated by the pre-trained Densenet121 model on the NCHLT dataset.}
\end{figure}

\begin{figure}[H]
    \centering
    \includegraphics[scale=0.8]{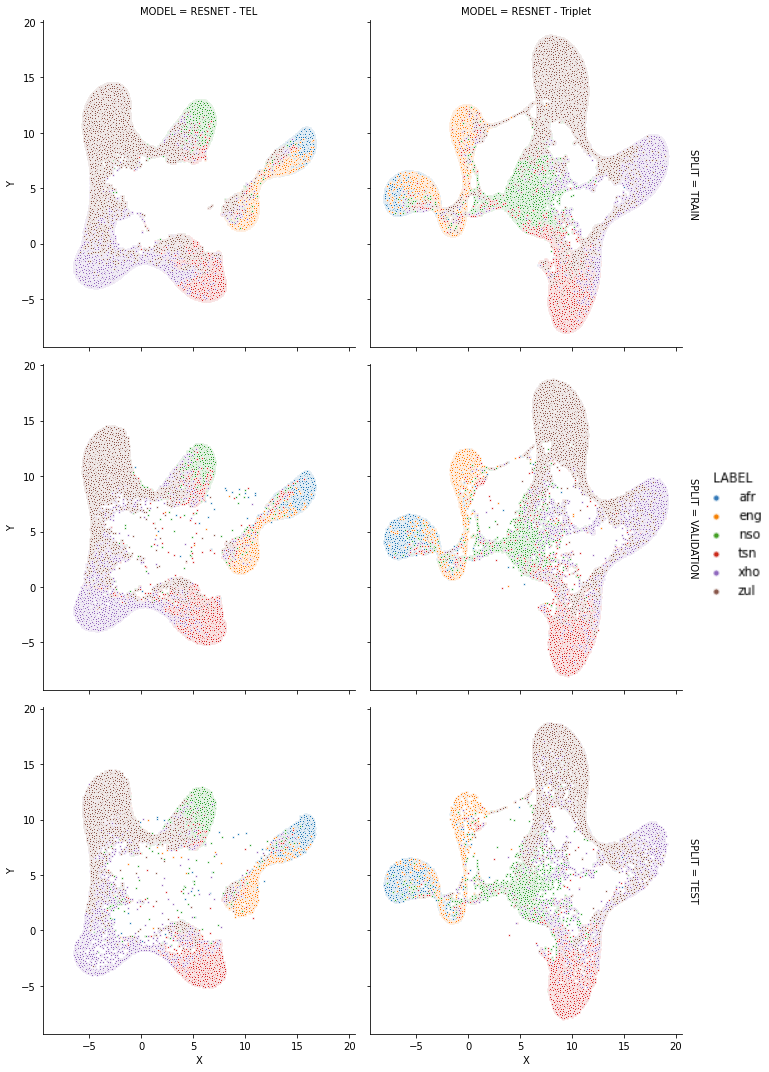}
    \caption{Embeddings generated by the baseline Resnet50 model on the NCHLT dataset.}
\end{figure}

\begin{figure}[H]
    \centering
    \includegraphics[scale=0.8]{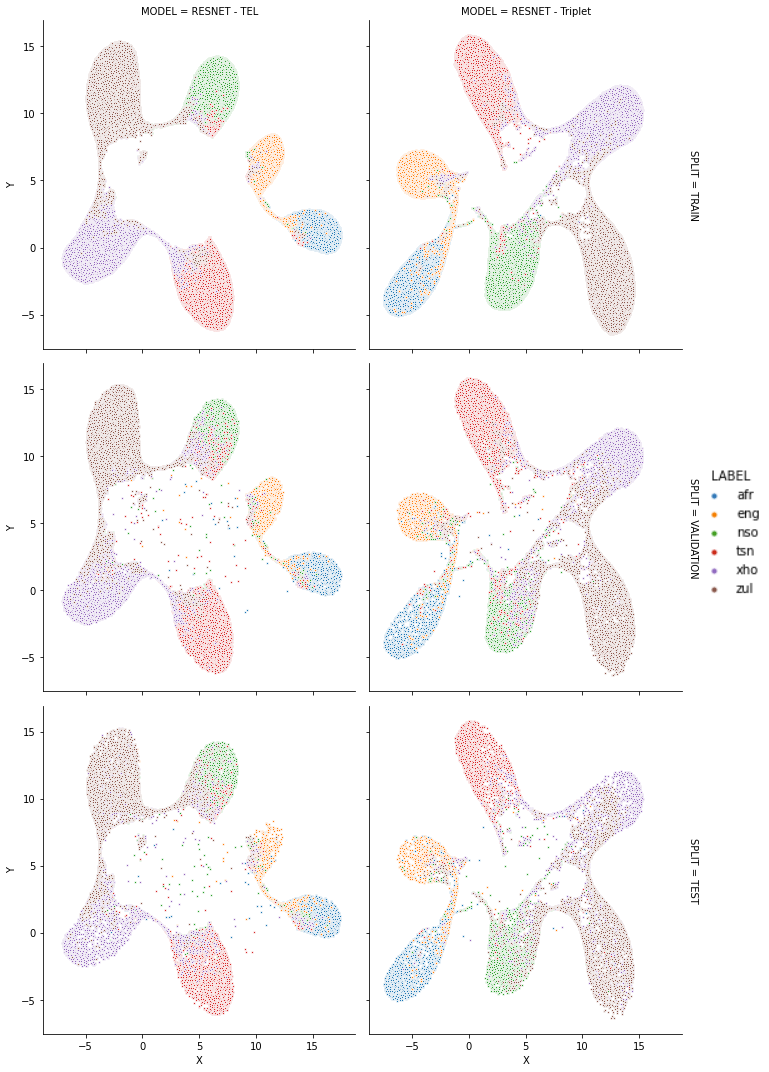}
    \caption{Embeddings generated by the Resnet50 model on the NCHLT dataset.}
\end{figure}

\begin{figure}[H]
    \centering
    \includegraphics[scale=0.8]{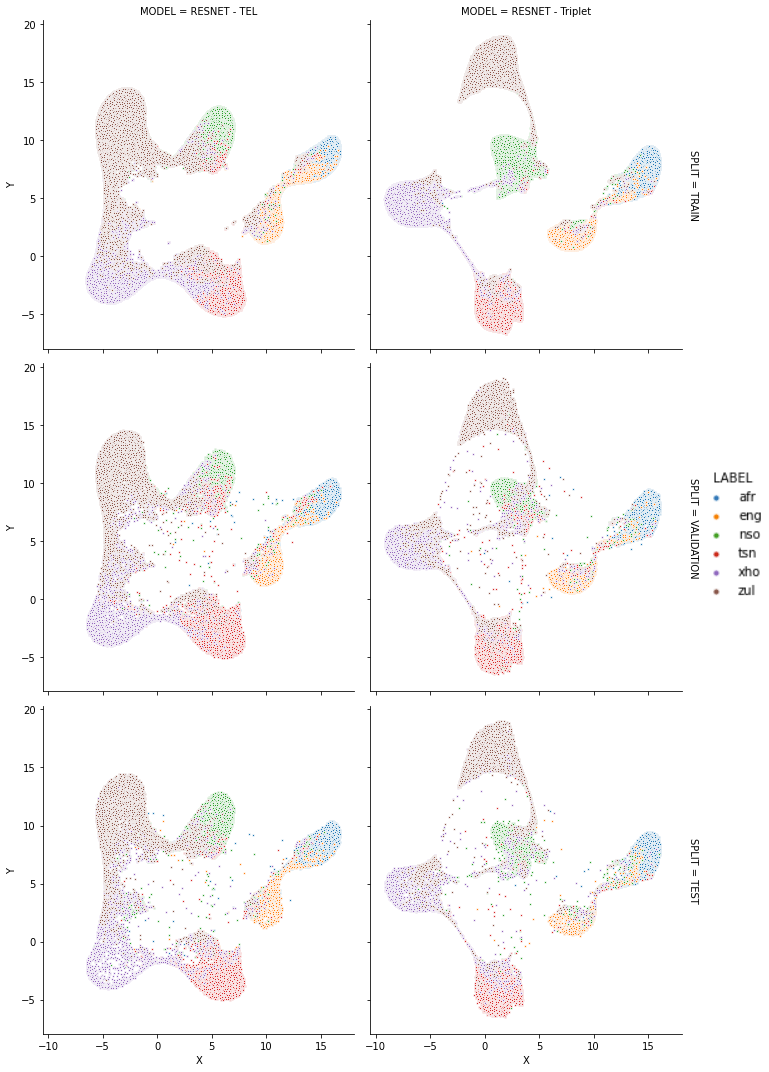}
    \caption{Embeddings generated by the baseline pre-trained Resnet50 model on the NCHLT dataset.}
\end{figure}

\begin{figure}[H]
    \centering
    \includegraphics[scale=0.8]{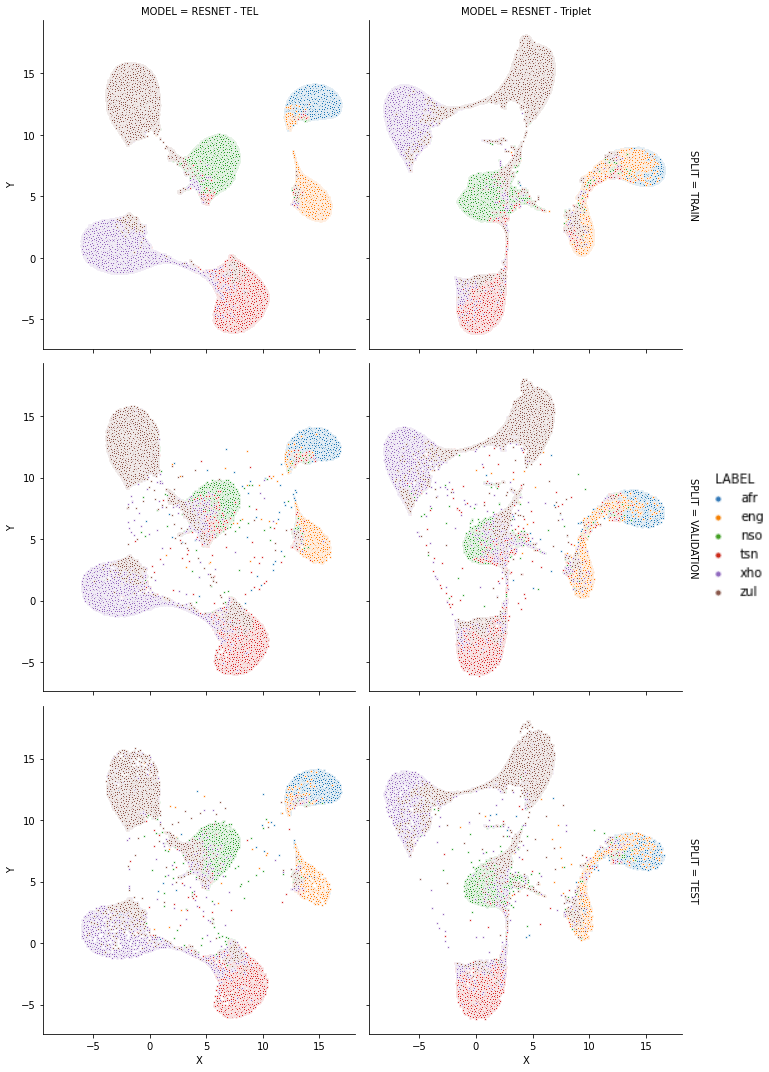}
    \caption{Embeddings generated by the  pre-trained Resnet50 model on the NCHLT dataset.}
\end{figure}

\end{document}